\documentclass[12pt]{iopart}
\usepackage{graphicx}
\usepackage{iopams}
 
\begin{document}

\title{Chaplygin gas in decelerating DGP gravity}

\author{Matts Roos}

\address{Department of Physical Sciences and Department of
          Astronomy\\ FIN-00014 University of Helsinki, Finland}
\ead{matts.roos@helsinki.fi}
\begin{abstract}
Explanations to the accelerated expansion of the Universe are
usually sought either in modifications of Einstein gravity or in new
forms of energy density. An example of modified gravity is the
braneworld Dvali-Gabadadze-Porrati (DGP) model which is
characterized by a length scale which marks the cross-over between
physics occurring in our four-dimensional brane and in a
five-dimensional bulk space. An example of dark energy is Chaplygin
gas which has similar asymptotic properties at early and late cosmic
times. Since Chaplygin gas gives too much acceleration we combine it
with the self-decelerating branch of the DGP model, taking the
cross-over scales to be proportional. This 3-parameter model fits
supernovae data with a goodness-of-fit equalling that of the
$\Lambda$CDM model. In contrast to generalized DGP models and
Chaplygin gas models, this model is unique in the sense that it does
not reduce to $\Lambda$CDM for any choice of parameters.
\end{abstract}
\maketitle

\section{Introduction}
The observation that the Universe is undergoing an
accelerated expansion has stimulated a vigorous search for models to
explain this unexpected fact. In General Relativity the Einstein tensor $G_{\mu\nu}$ encodes the geometry of the Universe, the stress-energy tensor $T_{\mu\nu}$ the energy density. Thus modifications to $G_{\mu\nu}$ imply  alternative geometries, modifications to $T_{\mu\nu}$ involve new forms of energy densities, called dark energy. The traditional solution is the cosmological constant $\Lambda$ which can be interpreted either as a modification of the geometry or as a vacuum energy term in $T_{\mu\nu}$. This fits observational data well, in fact no competing model does better.

A simple and well-studied model of modified gravity is the
Dvali--Gabadadze--Porrati (DGP) braneworld model (Dvali \& al.\,\cite{Dvali},
Deffayet \& al.\,\cite{Deffayet}) in which our four-dimensional world is an
FRW brane embedded in a five-dimensional Minkowski bulk. The model
is characterized by a cross-over length scale  $r_c$ such that
gravity is a four-dimensional theory at scales $r\ll r_c$ where
matter behaves as pressureless dust. In the self-accelerating branch, gravity "leaks out" into the bulk at scales $r \gg r_c$ and the model approaches the behavior of a cosmological constant. To explain the accelerated expansion which is of recent date ($z\approx
0.5$), $r_c$ must be of the order of $H_0^{-1}$. In the
self-decelerating DGP branch, gravity "leaks in" from the bulk at
scales $r\gg r_c$. Note that the self-accelerating branch has a ghost, whereas the self-decelerating branch is ghost-free.

A simple and well-studied model of dark energy introduces into $T_{\mu\nu}$ the density
$\rho_{\varphi}$ and pressure $p_{\varphi}$ of an ideal fluid called
Chaplygin gas \,\cite{Kamenshchik,Bilic}.
Like the previous model it is also characterized by a cross-over length scale
below which the gas behaves as pressureless dust, at late times approaching the behavior of a cosmological constant.

Since the standard 2-parametric Chaplygin gas model causes too much acceleration, we propose to combine it with the standard 2-parametric {\em self-decelerating} DGP model, taking the cross-over length scales to be proportional. The proportionality constant subsequently disappears because of a normalizing condition at $z=0$. Thus the model has only one parameter more than the standard $\Lambda$CDM model.

Generalizations of the Chaplygin gas model and the DGP model also have at least one parameter more than $\Lambda$CDM, yet they fit data best in the limit where they reduce to
$\Lambda$CDM. This model is a unique alternative in the
sense that it does not reduce to the $\Lambda$CDM model for any choice of parameters.

\section{Cross-over length scales}

On the four-dimensional brane in the DGP model, the action
of gravity is proportional to $M^2_{\rm Pl}$ whereas in the bulk it is
proportional to the corresponding quantity in 5 dimensions, $M^3_5$.
The cross-over length scale is defined~\cite{Deffayet} as
$r_c=M^2_{\rm Pl}/2M^3_5$, and an effective density parameter as
$\Omega_{r_c}=(2r_c H_0)^{-2}$. The Friedmann--Lema\^{\i}tre equation in the DGP model may be written~\cite{Deffayet}
\begin{equation}
H^2-\frac k{a^2}-\epsilon\frac 1 r_c\sqrt{H^2-\frac
k{a^2}}=\kappa\rho~,\label{Friedm}
\end{equation}
where $a=(1+z)^{-1},\ \kappa=8\pi G/3$, and $\rho$ is the total
cosmic fluid energy density $\rho$ with components $\rho_m$ for
baryonic and dark matter, and $\rho_{\varphi}$ for whatever additional dark energy may be present, in our case the Chaplygin gas. Clearly the standard FLRW cosmology is
recovered in the limit $r_c\rightarrow\infty$. In the following we shall only
consider $k=0$ flat-space geometry. The \emph{self-accelerating
branch} corresponds to $\epsilon=+1$, we shall consider the \emph{self-decelerating
branch} with $\epsilon=-1$.
The free parameters in the DGP model are then $\Omega_{r_c}$ and
$\Omega_m=\kappa\rho_m/H_0^2$.

The Chaplygin gas has the barotropic equation of state
$p_{\varphi}=-A/\rho_{\varphi}$ \cite{Kamenshchik,Bilic}, where
$A$ is a constant with the dimensions of energy density squared. The
continuity equation, $
\dot\rho_{\varphi}= -3H(\rho_{\varphi}-A/\rho_{\varphi})$, integrates to
\begin{equation}
\rho_{\varphi}(a)=\sqrt{A+B/a^6}~,\label{rhoch}
\end{equation}
where $B$ is an integration constant. The limiting behavior of the energy density is
\begin{equation}
\rho_{\varphi}(a)\propto\sqrt{B}/a^{3}~~ {\rm for}~~ a\approx(B/A)^{1/6},~\rho_{\varphi}(a)\propto \sqrt{A}~~  {\rm for}~~ a\gg (B/A)^{1/6}.
\end{equation}
We choose the two scales, $r_c$ and $(B/A)^{1/6}$, to be proportional by a factor $H^c_0$, so that
\begin{equation}
\left(\frac B A\right)^{1/6}=r_c H^c_0=\frac{H^c_0}{2H_0\sqrt{\Omega_{r_c}}}\ .\label{BA}
\end{equation}
The notation $H^c_0$ is chosen because this parameter has the dimension of $H_0$. Note however, that $H_0\equiv H(1)$ is a constant in Eq.~(\ref{Friedm}) (as well as in the standard model), whereas $H^c_0$ is a function of the parameters entering the combined model.

It is now convenient to replace the parameters $A$ and $B$ in Eq.\,(\ref{rhoch}) by $\Omega_A=\kappa\sqrt{A}/H^2_0$ and
$H^c_0=H_0\sqrt{4\Omega_{r_c}}(B/A)^{1/6}$. The dark energy density can then be written
\begin{equation}
\rho_{\varphi}(a)=H_0^2\kappa^{-1}\Omega_A \sqrt{1+(H^c_0/H_0)^6(4\Omega_{r_c}a^2)~^{-3}}~.\label{rhophi}
\end{equation}

\section{The combined model}

In the DGP model the expansion history $H(a)$ is given by Eq.\,(\ref{Friedm}). To distinguish our combined model (in which $\epsilon=-1$), we denote the expansion history $H^c(a)$. Making use of the definitions of $\Omega_{r_c}$, $\Omega_m=\Omega_m^0/a^3$, $\Omega_A$, $H^c_0$, and $\rho_{\varphi}(a)$, the Friedmann equation becomes
\begin{equation}
\frac {H^c(a)}{H_0}=-\sqrt{\Omega_{r_c}}+\left[\Omega_{r_c}+\Omega_m^0 a^{-3}
+\Omega_A\sqrt{1+(H^c_0/H_0)^6(4\Omega_{r_c}a^2)~^{-3}}\right]^{1/2}.\label{Ha}
\end{equation}
Note that $\Omega_{r_c}$ and $\Omega_A$ do not evolve with $a$, just like $\Omega_{\Lambda}$ in the the $\Lambda$CDM model.

A closer inspection of Eq.~(\ref{Ha}) reveals that it is not properly normalized at $a=1$  to a constant ${H^c(1)}/{H_0}$, because the right-hand-side takes different values at different points in the space of the parameters $\Omega_m^0,~\Omega_{r_c},\Omega_A$, and $H^c_0$. This gives us a condition: at $a=1$ we require that $H^c(1)=H^c_0$.

This condition is a
6:th order algebraic equation in the variable $H^c_0/{H_0}$. Finding the real, positive roots and inserting the smallest one in Eq.~(\ref{Ha}) normalizes the equation properly, and rids us of the parameter $H^c_0$, which henceforth is a function $H^c_0=f(\Omega_m^0,~\Omega_{r_c},\Omega_A)$. The only problem is that the function cannot be expressed in closed form, so one has to resort to numerical iterations.

In this brief presentation we shall make the approximation
$H^c(1)=H_0$ which is very good, except for at $z\ll 0.1$, leaving
the general solution to a later paper (in preparation). The correct
value of  $H^c_0/{H_0}$ is about 0.96.

\section{Data, analysis, and fits}
The data we use to test this model are the same 192 SNeIa data as in the compilation used by Davis \& al.\,\cite{Davis} which is a combination of the "passed" set in Table 9 of Wood-Vasey \& al.\,\cite{Wood} and the "Gold" set in Table 6 of Riess \& al.\,\cite{Riess}.

We are sceptical about using CMB and BAO power spectra, because they
have been derived in FRW geometry, not in five-dimensional brane geometry.
The SNeIa data are, however, robust in
our analysis, since the distance moduli are derived from light
curve shapes and fluxes, that do not depend on the choice of
cosmological models.

We compute luminosity distances and magnitudes for all the supernovae using the expression (\ref{Ha}). The $\chi^2$ sum turns out to be very insensitive
to values of $\Omega_m^0$  and $\Omega_A$ far from the best fit, a problem we cure by adding a weak CMB prior, $\Omega_m^0=0.239\pm 0.09$.

The calculations are done with the classical CERN program MINUIT (James \& Roos\,\cite{James}) which delivers $\chi^2_{\,best}$, parameter errors, error contours and parameter correlations. We do not marginalize, but quote the full, simultaneous
confidence region: a $1\sigma$ error contour corresponds to $\chi^2_{\,best} + 3.54$. The best fit is at the parameter values
\begin{equation}
\Omega_m^0=0.26\pm 0.16,~\Omega_{r_c}=0.82^{+0.69}_{-0.22},~\Omega_A=2.21^{+0.50}_{-0.22}\label{res}
\end{equation}
with $\chi^2=195.5$ for $193-3$ \emph{d.f.} ($\chi^2/d.f.=1.029$), exactly the goodness-of-fit of the $\Lambda CDM$ model for the same data.

In Fig.\,1 we plot the best fit confidence region in the
$\Omega_{r_c}$ vs $\Omega_m^0$-plane, a banana-shaped closed contour. A cross in the Figure marks the point of best fit (\ref{res}).

A further constraint is obtained from U/r chronometry by Frebel \& al.~\cite{Frebel} of the age of the oldest star HE 1523-0901, $t_* = 13.4\pm 0.8\,(stat)\pm 1.8\,(U~ production~ ratio)$.  This can be written as a lower limit to the age of the Universe $t(Universe) > 12$ Gyr (68\%C.L.). The resulting forbidden region is blue in Fig.\,1. The best point obtained from the supernova fit is now in the blue region, thus this age condition has a real influence on the allowed parameter space.

\section{Conclusions}
We embed standard Chaplygin gas in the ghost-free self-decelerating DGP geometry with an extra condition: that the cross-over length scales, $r_c$ and $(B/A)^{1/6}$ respectively, are proportional. The proportionality factor is fixed by a normalizing condition at $z=0$, leaving three free parameters to be adjusted -- one more than in the standard $\Lambda$CDM model. The resulting cosmic acceleration fits supernova data~\cite{Davis,Wood,Riess} excellently. It can also be constrained by the distance to the Last Scattering Surface, and by a stellar lower limit to the age of the Universe \cite{Frebel}.

Generalizations of the Chaplygin gas model and the DGP model also have at least one parameter more than $\Lambda$CDM, yet they fit data best in the limit where they reduce to
$\Lambda$CDM. This model is a unique alternative in the
sense that it does not reduce to the $\Lambda$CDM model for any choice of parameters.

The effective equation-of-state changed from super-acceleration to acceleration sometime in the range $0 < z < 1$, approaching $w_{eff}=-1$ in the future. The 'coincidence problem' is a consequence of the time-independent value of $r_c$, a braneworld property.

 \begin{figure}[htbp]
\includegraphics{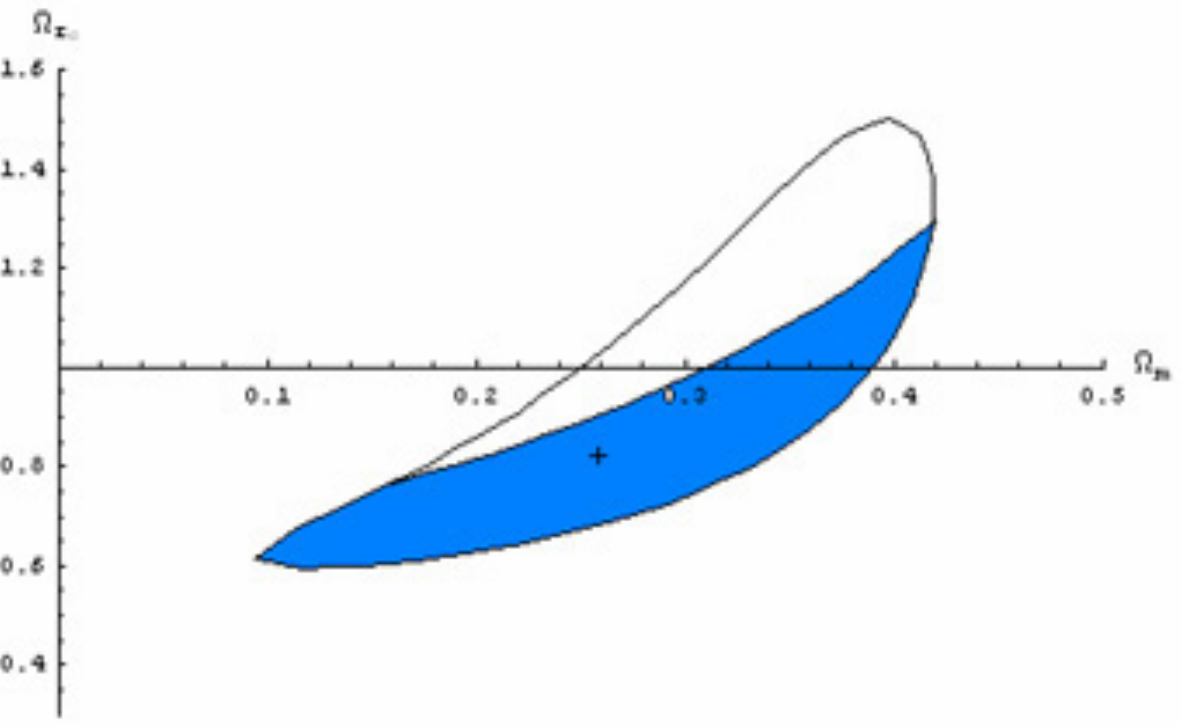}
   \caption{The closed contour is the confidence region in the $(\Omega_m^0,\Omega_{r_c})$-plane from a fit to SNeIa data. The point of best fit is marked by a cross. The blue region is forbidden because it would make th Universe too young.}
     \end{figure}

\end{document}